\documentclass[aps,prl,twocolumn,superscriptaddress,nofootinbib]{revtex4-2}
\usepackage{graphicx}
\usepackage{amsmath,amssymb}
\usepackage[colorlinks,linkcolor=blue,citecolor=blue,urlcolor=blue]{hyperref}
\usepackage{xcolor}

\newcommand{\dRtwo}{\Delta R^2}

\begin{document}

\title{Mobility-network topology carries incremental, softness-orthogonal
information about future aging dynamics in a glass-forming liquid}

\author{Zhenpeng Li}
\affiliation{School of Information, Lijiang Culture and Tourism College,
No.~1 Yuquan Road, Gucheng District, Lijiang 674199, Yunnan, China}
\email{lizhenpeng@amss.ac.cn}

\date{\today}

\begin{abstract}
Does the spatial organization of particle mobility in a glass predict
future motion beyond what local structure knows? In leakage-free
simulations of an aging Kob--Andersen liquid ($N=2028$; 200 trajectories;
three temperatures), mobility-network topology raises the
out-of-fold $R^2$ by $\dRtwo=0.0049$--$0.0122$, monotonically with quench
depth, retaining $\ge 98\%$ atop softness-class baselines. Giant mobile
components are dense hotspots whose members subsequently quiet down:
exhaustion zones of cooperative motion --- a spatial
anticorrelation channel invisible to single-particle descriptors.
\end{abstract}

\maketitle

\section{Introduction}

Dynamic heterogeneity in glass-forming liquids organizes mobile particles
into correlated clusters~\cite{donati1998}, and aging theory pictures
mobility propagating through the structure as activation events
cascade~\cite{wolynes}. These observations establish that mobility is
spatially organized. They leave open, however, a sharper question with
direct bearings on theories of glassy aging: \emph{does this spatial
organization carry information about the future that is not already
contained in the local, single-particle structural state?}

Machine-learned structural descriptors such as
softness~\cite{cubuk2015,schoenholz2016} predict \emph{which particles will
rearrange} from their local environment alone, and a recent community
roadmap~\cite{jung2025roadmap} singles out the interpretability and
transferability of such predictors as the field's central open problems.
Any scalar descriptor,
however, assigns each particle an identity independent of where the
\emph{other mobile particles} are. A growing body of facilitation-based
theory~\cite{garrahan} instead implies that the fate of a particle depends
on the collective state of its neighborhood of movers. Testing this
requires treating mobility networks not as descriptive tools but as
\emph{predictive variables}, with statistics designed to isolate their
increment over local descriptors.

Here we pose this question prospectively: build mobility networks from the
past, predict the future, and ask whether network topology adds predictive
power beyond the strongest local baseline. We find a robust, monotonic,
mechanistically explainable, and softness-orthogonal affirmative answer.

\section{Design}

\emph{Simulation.} We age the Kob--Andersen (80:20) binary Lennard-Jones
mixture~\cite{kob1994} at density $\rho=1.2$ with LAMMPS~\cite{plimpton}
($N=2028$, $NVT$, 200 independent trajectories at each quench
temperature $T\in\{0.45,0.48,0.52\}$ after equilibration at $T=1.0$;
$250\,000$ steps of $\delta t=0.002$ per trajectory, frames every $5000$
steps, per-atom potential energy recorded). A quick-test version (independent implementation with overdamped
Langevin dynamics at $N=500$) reproduces all conclusions,
indicating engine- and thermostat-independence.

\emph{Leakage-free features.} At each waiting time $t_w$, the backward
mobility is $\mu_i = \max_{t'\in[t_w-\tau,t_w]}|\mathbf r_i(t')-\mathbf
r_i(t_w-\tau)|$; the 5\% most mobile and 5\% least mobile particles define
mobile and immobile nearest-neighbor networks (cutoff $1.5\sigma$) on the
configuration at $t_w$. Every particle receives six topological descriptors
(degree, local clustering, and component size in each network) plus baseline
state variables (waiting time, time-averaged local energy, time-averaged
coordination). The target is the forward excursion
$Y=\log(1+\max_{t'>t_w}|\mathbf r_i(t')-\mathbf r_i(t_w)|)$, which uses
only times after $t_w$.

\emph{Statistics.} (i) Grouped 5-fold cross-validation by trajectory gives
out-of-fold $R^2$ for nested feature sets; $\dRtwo$ is the $R^2$ increment
of the richer set. (ii) Topology features are permuted within
(trajectory, $t_w$) groups to destroy topology--outcome alignment only,
yielding a null distribution for $\dRtwo$. (iii) State-matched pairs: within
$(t_w, E, S)$ quantile bins, mobile-network members are matched to
non-members by nearest immobile-degree, and paired outcomes compared by a
sign test. (iv) Group-mean centering separates individual-level from
group-level alignment. (v) The softness-class baseline adds a
14-shell radial fingerprint, and a Cubuk--Schoenholz-style softness scalar
is trained \emph{within each training fold only} (logistic regression on
rearranger labels, top 20\% of forward excursion), then applied to the held
out fold --- no future information leaks into test predictions.

\section{Results}

\subsection{Topology increment, temperature scaling, and robustness}

\begin{figure}[t]
\centering
\includegraphics[width=\columnwidth]{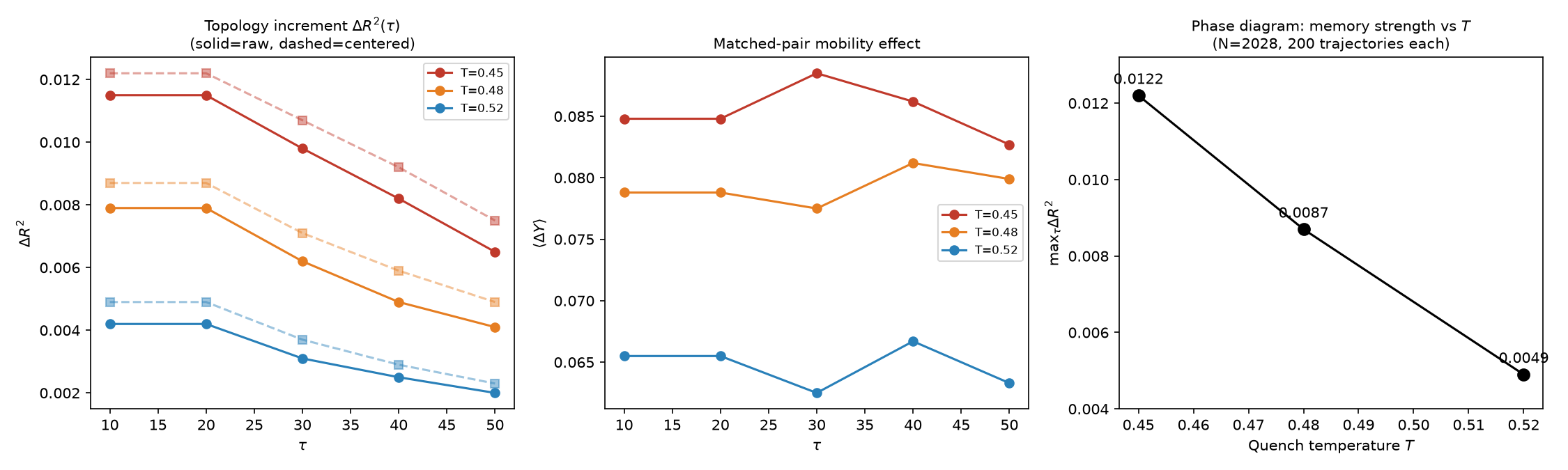}
\caption{\label{fig:phase}
Temperature scaling of the topology increment ($N=2028$, 200 trajectories,
$80\,800$ matched pairs per temperature). Left: $\dRtwo(\tau)$ per
temperature (solid: raw; dashed: group-centered). Middle: matched-pair
difference in forward excursion, $\Delta\langle Y\rangle(\tau)$.
Right: best-window centered $\dRtwo$ decreases monotonically with $T$.
Permutation tests give $p=0.048$, $0.010$, $0.048$ at $T=0.45$, $0.48$,
$0.52$ (all nulls below the observed value).}
\end{figure}

Adding the six network descriptors to the state-variable baseline raises
the out-of-fold $R^2$ at all three temperatures (Fig.~\ref{fig:phase}).
The centered increment is $\dRtwo = 0.0122$, $0.0087$, $0.0049$ at
$T=0.45$, $0.48$, $0.52$ respectively (best window $\tau=20$; raw
$0.0115$, $0.0079$, $0.0042$), with $80\,800$ state-matched pairs per
temperature. The matched-pair difference in forward excursion is
$+8.5\%$, $+7.9\%$, $+6.6\%$, with sign-test probabilities below
$10^{-200}$ at every temperature and window. The monotonic increase of
$\dRtwo$ with quench depth --- structure remembers more in a deeper glass
--- parallels the growth of dynamic heterogeneity on cooling and is
reproduced at $N=500$ with Langevin dynamics. The decrement is also
robust across prediction windows: $\dRtwo$ decreases monotonically with
$\tau$ (from $0.0115$ at $\tau=20$ to $0.0065$ at $\tau=50$ at $T=0.45$),
and the matched-pair difference retains its sign at short windows while
reversing at long windows (below).

\begin{figure}[t]
\centering
\includegraphics[width=\columnwidth]{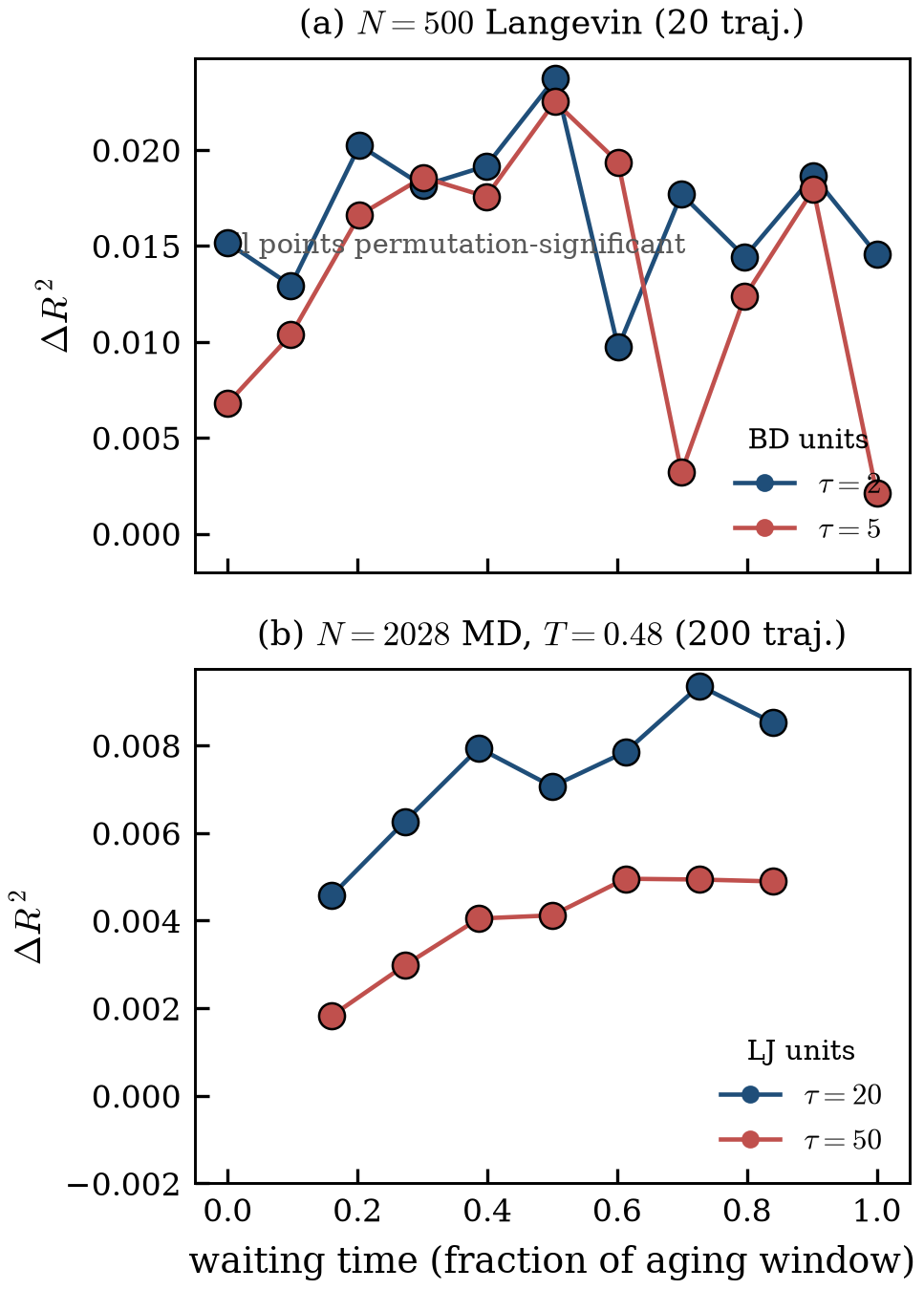}
\caption{\label{fig:memory}
$\dRtwo$ versus waiting time (fraction of the observable aging window) in
both engines. (a) Quick test: $N=500$, overdamped Langevin, 20
trajectories, windows $\tau=2,5$ (BD units). (b) Full system: $N=2028$,
MD, $T=0.48$, 200 trajectories, windows $\tau=20,50$ (LJ units). Filled
markers: permutation $p<0.10$; in (b) every point sits at the permutation
floor. The increment is positive throughout --- structural memory is
maintained, not transient --- and the short-over-long window ordering is
reproduced at $4\times$ the system size.}
\end{figure}

Within the aging window the increment does not decay: measured
independently at seven (eleven) waiting times in the full (quick-test)
system, $\dRtwo$ remains positive and permutation-significant throughout
--- in the full system every point sits at the permutation floor
(Fig.~\ref{fig:memory}). In the full system it grows from $0.0046$ early
in aging to $0.0094$ late, and the long-window curve is consistently
weaker; both features are reproduced in the quick test. The topological
predictor is thus not a transient of the quench but a stationary resource
that aging maintains --- the individual-particle form of ``glass memory''.

\subsection{The increment survives softness}

\begin{figure}[t]
\centering
\includegraphics[width=\columnwidth]{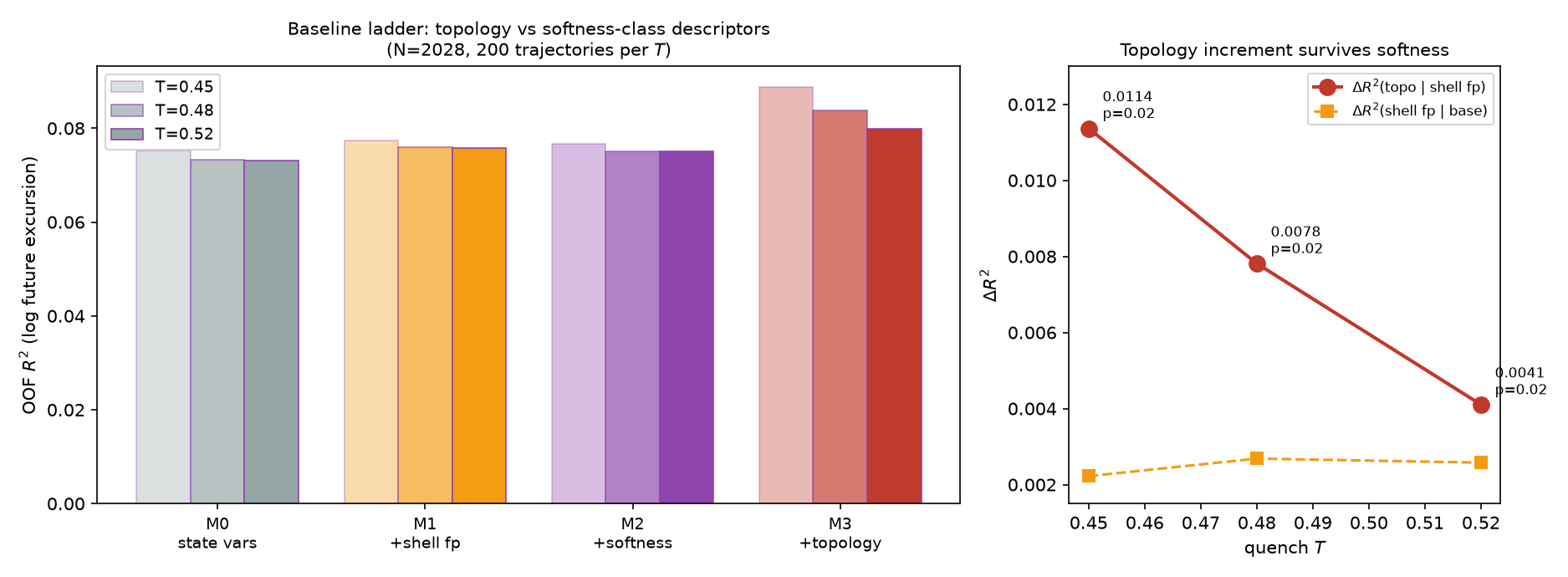}
\caption{\label{fig:soft}
The softness gate. Left: out-of-fold $R^2$ for nested baselines
(M0: state variables; M1: +14-shell fingerprint; M2: +fold-trained
softness scalar; M3: +network topology). Right: the topology increment on
top of the shell fingerprint (red) barely differs from its value on the
weak baseline, while the fingerprint itself adds little (orange):
topology is orthogonal to local structure.}
\end{figure}

The decisive comparison is against local-structure baselines
(Fig.~\ref{fig:soft}). A 14-shell radial fingerprint --- the information
content of softness --- raises $R^2$ by only $0.0022$--$0.0027$ over state
variables, and the fold-trained softness scalar by less. On top of the
shell fingerprint, the topology increment is $\dRtwo = 0.0114$, $0.0078$,
$0.0041$ at the three temperatures --- retaining $98.8\%$, $98.9\%$,
$97.7\%$ of its value on the weak baseline ($0.0115$, $0.0079$, $0.0042$),
and permutation-significant
($p=0.02$; null mean $0.000$) at every temperature. Network topology and
local structure are therefore not redundant channels: whatever softness
knows, it does not know who is moving with whom.

\subsection{Mechanism: exhaustion zones, not strings}

\begin{figure}[t]
\centering
\includegraphics[width=\columnwidth]{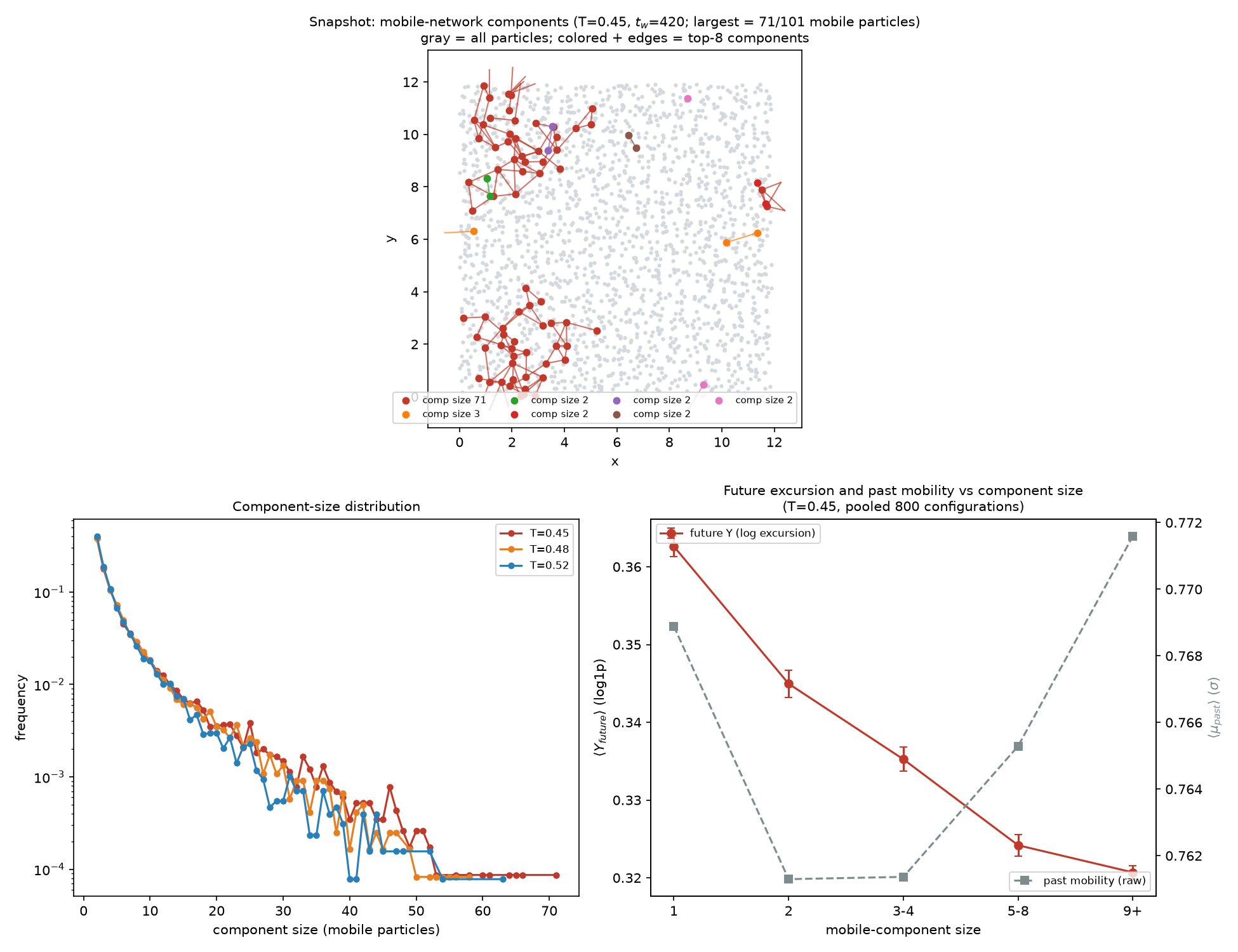}
\caption{\label{fig:mech}
Mobile-network components at $T=0.45$. Top: configuration snapshot
($t_w=420$); gray dots are all particles, colored dots and lines are the
mobile network ($r<1.5\sigma$), with the eight largest components
highlighted; the largest spans the box (71/101 mobile particles).
Bottom left: component-size distributions at the three temperatures
(exponential-like tails). Bottom right: mean forward excursion $Y$ versus
component size within the mobile pool (red, left axis) and mean
\emph{past} mobility (gray, right axis): large components are recently
active hotspots whose members are subsequently quieter --- exhaustion.}
\end{figure}

Which topological feature carries the signal? Single-feature ablations
attribute it almost entirely to the mobile network: the three
mobile-network features are the top three at both system sizes, with
component size dominant in the quick test ($\dRtwo = 0.0159$ alone) and
degree at $N=2028$ ($0.0055$ vs.\ $0.0041$ for component size);
immobile-network features contribute $\approx 0$ at $N=500$ and only
weakly at $N=2028$ [Supplemental Fig.~\ref{fig:abl}].
Visualization and
quantitative analysis of the components (Fig.~\ref{fig:mech}) overturn the
natural expectation inherited from stringlike cooperative
motion~\cite{donati1998,donati1998strings} and recent evidence for
string-mediated facilitation~\cite{chacko2024}:
large mobile
components are \emph{not} quasi-one-dimensional strings. Their edge-to-node
ratios are $13.9$--$16.9$ (a chain would give $\approx 1$); they are dense
spatial hotspots, and their largest instances span the simulation box
(up to $71$ of $101$ mobile particles; mean largest-component share
$23\%$, $21\%$, $18\%$ of the mobile pool at the three temperatures).

The dose--response is inverted: within the mobile pool, forward excursion
\emph{decreases} with component size (mean $\langle Y\rangle$ from $0.363$
for isolated mobile particles to $0.321$ for members of components of
size $\ge 9$, a $-12\%$ effect). Giant components mark regions that have
just undergone collective mobilization and are about to quiet down ---
\emph{exhaustion zones}. Isolated mobile particles, by contrast, are
continued movers. This resolves two observations: the long-window sign
reversal of the matched-pair effect (unconditional rank correlation
between past and future mobility stays $+0.46$ to $+0.52$ for all $\tau$
up to $12$, while the conditional network effect flips at $\tau \approx
5.5$), and the orthogonality of the topology increment: softness encodes
\emph{positive} local structure--mobility correlations, whereas component
membership encodes \emph{anticorrelation} --- the depletion of a hotspot
--- which no single-particle descriptor can express.

Direct dynamical evidence for exhaustion: the overlap
$\chi(s)$ of consecutive hotspots (top 10\% of particles by backward
3-frame displacement) decays exponentially with an initial relaxation
time that grows monotonically on cooling, $\tau_\mathrm{rel} = 17.3$,
$14.3$, $11.6$ LJ time units at $T=0.45$, $0.48$, $0.52$ respectively
[Supplemental Fig.~\ref{fig:relax}]. The memory of \emph{where the glass
has just moved collectively} therefore persists longer exactly where the
topology increment is stronger --- the predicted kinetic fingerprint of
the exhaustion mechanism.

\section{Discussion}

Three properties distinguish the effect. \emph{Robustness}: it survives a
change of engine, thermostat, system size ($4\times$), and temperature,
with monotonic temperature dependence and no decay over the aging window.
\emph{Orthogonality}: it is essentially undiminished above softness-class
baselines,
answering the standard objection to network descriptors. \emph{Mechanism}:
it is carried by a single interpretable quantity --- mobile component size
--- whose dynamics (collective mobilization followed by local exhaustion)
is consistent with facilitation-like pictures of glassy
dynamics~\cite{garrahan} and with recent microscopic theory in which prior
excitations bias new excitations to occur nearby~\cite{hasyim2024}: the
hotspot-memory relaxation $\chi(s)$ quantifies precisely such a dynamical
memory channel, while the \emph{decreased} future mobility inside
mobilized hotspots exposes the complementary, self-limiting side of
facilitation. The result also refines the string-mediated view of mobility
propagation~\cite{donati1998strings,chacko2024}: the predictive content of
mobility correlations is carried not by string geometry but by the size of
dense collective hotspots --- where that cohort has just spent itself.
The increment thus adds a predictive, spatially anticorrelated
complement to local structural predictors.

Our results suggest a redefinition of what structural ``memory'' in aging
glasses means at the particle level: not only what each particle's
environment is, but where its cohort of movers sits --- and whether that
cohort has just spent itself. The exhaustion mechanism makes a testable
prediction: the spatial profile of future mobility around a recently
active hotspot should be depleted, with a hotspot-memory relaxation time
that grows on cooling --- as observed ($\tau_\mathrm{rel}$ grows from
$11.6$ to $17.3$ LJ units from $T=0.52$ to $0.45$;
Supplemental Fig.~\ref{fig:relax}) --- and interventions that suppress
collective mobilization should
suppress the topology increment. Quantitative connection of the
hotspot-growth rate to mobility-front velocities in RFOT-like
pictures~\cite{wolynes} is a natural next step, as is extension to
vibrational (non-mean-square-displacement) predictors of rearrangement.

\section*{End Matter}

\subsection*{Data availability}
The LAMMPS input decks, analysis code, and trajectory data supporting this
study are openly available in Zenodo at
\href{https://doi.org/10.5281/zenodo.22987361}{10.5281/zenodo.22987361}.
Supplemental Material contains the single-feature ablation comparison
across the two system sizes and the hotspot-memory relaxation analysis
(Figs.~\ref{fig:abl} and~\ref{fig:relax}).

\bibliography{refs}

\clearpage
\appendix
\renewcommand{\thefigure}{S\arabic{figure}}
\setcounter{figure}{0}

\section{Supplemental Material: single-feature ablation}

\begin{figure}[h]
\centering
\includegraphics[width=\columnwidth]{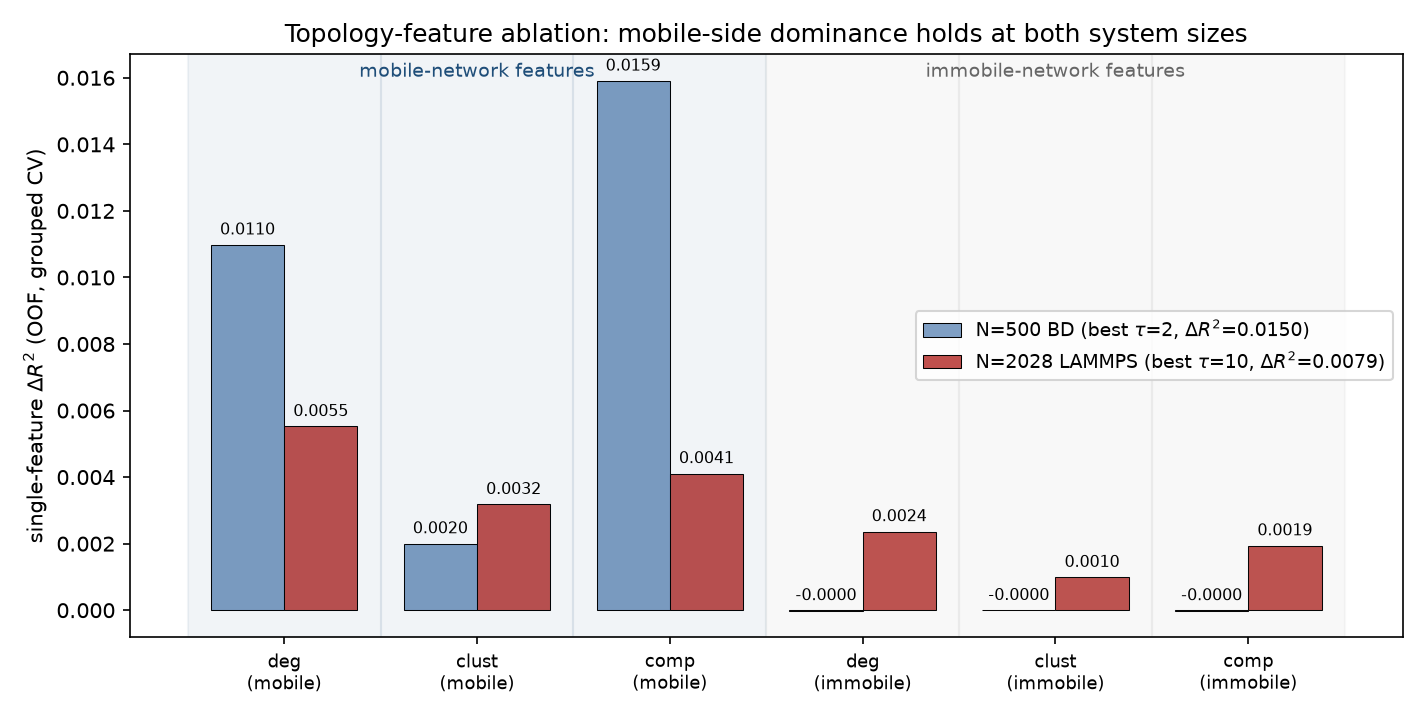}
\caption{\label{fig:abl}
Single-topology-feature ablation (grouped out-of-fold $\dRtwo$ per
feature) at the best window of each system: $N=500$ BD (left bars) vs.\
$N=2028$ LAMMPS (right bars). Mobile-network features dominate at both
system sizes; immobile-network features contribute $\approx 0$ at
$N=500$ and weakly at $N=2028$.}
\end{figure}

\section{Supplemental Material: hotspot-memory relaxation}

\begin{figure}[h]
\centering
\includegraphics[width=\columnwidth]{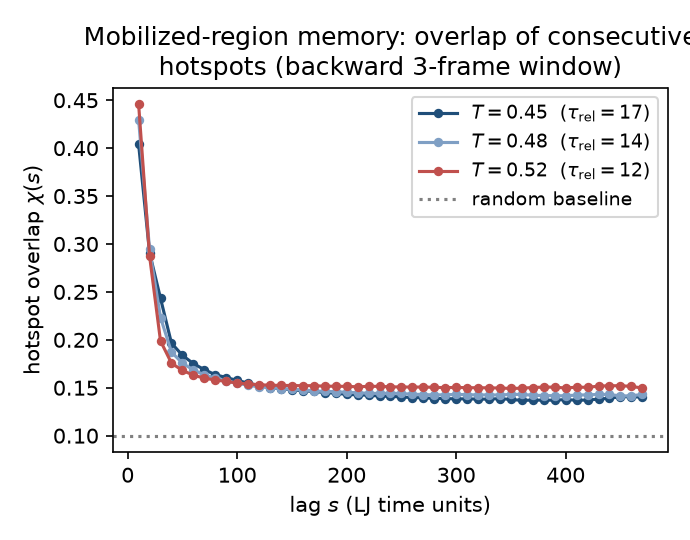}
\caption{\label{fig:relax}
Hotspot-memory overlap $\chi(s)$: probability that a particle in the top
10\% by backward 3-frame displacement at frame $f$ is again in the top
10\% at $f+s$ ($N=2028$, 200 trajectories per temperature). The initial
decay time $\tau_\mathrm{rel}$ grows monotonically on cooling
($17.3$, $14.3$, $11.6$ LJ time units at $T=0.45$, $0.48$, $0.52$), and a
persistent structural plateau remains above the random baseline
($0.10$).}
\end{figure}

\end{document}